\documentclass[useAMS,usenatbib,usegraphicx]{mn2e}
\onecolumn{}

\title[KIC 6206751: a pulsating R CMa eclipsing binary]{KIC 6206751: the first R CMa-type eclipsing binary with $\gamma$ Doradus pulsations}

\author[Lee \& Park]
       {Jae Woo Lee$^{1,2}$\thanks{E-mail: jwlee@kasi.re.kr} and Jang-Ho Park$^{1,3}$ \\
        $^1$Korea Astronomy and Space Science Institute, Daejeon 34055, Korea \\
        $^2$Astronomy and Space Science Major, Korea University of Science and Technology, Daejeon 34113, Korea \\
        $^3$Department of Astronomy and Space Science, Chungbuk National University, Cheongju 28644, Korea }

\begin{document}

\date{Accepted 2018 ---------. Received 2018 ---------; in original form 2018 }

\pagerange{\pageref{firstpage}--\pageref{lastpage}} \pubyear{2018}

\maketitle

\label{firstpage}

\begin{abstract}
We present the absolute properties of the double-lined eclipsing binary KIC 6206751 exhibiting multiperiodic pulsations. 
The ${\it Kepler}$ light curve of this system was simultaneously solved with the radial-velocity data given by 
Matson et al. (2017). The results indicate that the binary star is a short-period semi-detached system with 
fundamental parameters of $M_1$ = 1.66$\pm$0.04 M$_\odot$, $M_2$ = 0.215$\pm$0.006 M$_\odot$, $R_1$ = 1.53$\pm$0.02 R$_\odot$, 
$R_2$ = 1.33$\pm$0.02 R$_\odot$, $L_1$ = 5.0$\pm$0.6 L$_\odot$, and $L_2$ = 0.96$\pm$ 0.09 L$_\odot$. We applied 
multiple frequency analyses to the eclipse-subtracted light residuals and detected the 42 frequencies below 2.5 d$^{-1}$. 
Among these, three independent frequencies of $f_2$, $f_3$, and $f_4$ can be identified as high-order (38 $\le n \le$ 40) 
low-degree ($\ell$ = 2) gravity-mode oscillations, whereas the other frequencies may be orbital harmonics and 
combination terms. The ratios between the orbital frequency and the pulsation frequencies are 
$f_{\rm orb}$:$f_{\rm 2-4}$ $\simeq$ 2:3, which implies that the $\gamma$ Dor pulsations of the detached primary star 
may be excited by the tidal interaction of the secondary companion. The short orbital period, and the low mass ratio 
and $M_2$ demonstrate that KIC 6206751 is an R CMa-type star, which is most likely evolving into an EL CVn star. Of seven 
well-studied R CMa-type stars, our program target is the only eclipsing binary with a $\gamma$ Dor pulsating component. 
\end{abstract}


\section{INTRODUCTION}

R CMa-type stars are eclipse binaries (EBs) with characteristics of both a short orbital period and a low mass ratio 
among Algol systems (Varricatt \& Ashok 1999; Budding \& Butland 2011; Lehmann et al. 2013; Lee et al. 2016b). 
The initially more massive star becomes the present oversized secondary with a mass less than $\sim$0.3 M$_\odot$ through 
mass loss caused by stellar wind and mass transfer, and the gainer becomes the present early-type main-sequence primary as 
the result of mass accretion (Lee et al. 2018). The R CMa-type EBs are thought to have formed by non-conservative binary 
evolution and ultimately evolve into EL CVn stars, which are composed of a massive A(F)-type main-sequence star and 
a hotter helium white dwarf precursor with a mass of $\sim$0.2 M$_\odot$ in almost constant luminosity phase 
(Maxted et al. 2014; Chen et al. 2017). At present, there have only been six R CMa-type binaries with reliable physical 
parameters: R CMa (Budding \& Butland 2011; Lehmann et al. 2018), AS Eri (van Hamme \& Wilson 1984; Ibano\v{g}lu et al. 2006), 
OGLEGC 228 (Kaluzny et al. 2007), KIC 10661783 (Southworth et al. 2011; Lehmann et al. 2013), KIC 8262223 (Guo et al. 2017), 
and OO Dra (Zhang et al. 2014; Lee et al. 2018). Five stars except for OGLEGC 228 are EBs with a pulsating component, and 
they all exhibit $\delta$ Sct-type pulsations. 

Pulsating EBs that show both eclipses and pulsations serve as a Rosetta Stone for the study of stellar structure and 
evolution through asteroseismology and binary properties. About 22 of them have been known to contain $\gamma$ Dor-type 
pulsating components (Ibanoglu, \c Cakirli \& Sipahi 2018). $\gamma$ Dor pulsators are A$-$F stars of luminosity class 
IV$-$V pulsating in high-order gravity ($g$) modes with periods of 0.4$-$3 d and pulsation constants of $Q >$ 0.23 d 
(Kaye et al. 1999; Henry, Fekel \& Henry 2005). The pulsations are driven by a mechanism known as convective blocking 
(Guzik et al. 2000). Generally, the $\gamma$ Dor stars are cooler than $\delta$ Sct pulsators, which pulsate in 
low-order pressure ($p$) modes driven by the $\kappa$ mechanism with relatively short periods of 0.02$-$0.2 d and 
small pulsation constants of $Q <$ 0.04 d (Breger 2000). Nonetheless, the overlap in their instability regions 
indicates the possible existence of hybrid pulsators exhibiting two types of pulsations. Recently, 
the $\gamma$ Dor$-\delta$ Sct hybrids in several EBs have been detected from space missions such as {\it Kepler} and 
{\it CoRot} (see Samadi Ghadim, Lampens \& Jassur 2018). Such hybrid stars provide significant information about 
the structure from core to surface layers because the $g$ modes help to probe the deep interior near core region of 
a pulsator and the $p$ modes to probe its envelope (Kurtz et al. 2015). 

In order to advance this subject, we have been looking for pulsating components in EBs using the highly precise {\it Kepler} 
data (Lee et al. 2014, 2016a,b, 2017; Lee 2016) and then performing high-resolution time-series spectroscopic observations 
(Hong et al. 2015, 2017; Koo et al. 2016; Lee et al. 2018). This paper presents a continuation of detailed studies of pulsating EBs. 
KIC 6206751 (ASAS J192938+4130.8, TYC 3142-1295-1; $K_{\rm p}$ = $+$12.142, $V\rm_T$ = $+$12.527, $(B-V)\rm_T$ = $+$0.142) 
was announced as a short-period Algol with a binary period of about 1.2453 d (Hartman et al. 2004; Pigulski et al. 2009). 
Using the eclipse times from the {\it Kepler} photometry, Gies et al. (2012, 2015) reported that the orbital period of 
the system is decreasing. The authors assumed that the timing variation could be caused by the light-travel-time effect 
due to the possible presence of a circumbinary object. Gies et al. (2012) also noticed both pulsation and starspot activity 
in the light curve. On the other hand, double-lined radial-velocity (RV) curves of our program target were presented by 
Matson et al. (2017, hereafter MGGW). They derived the velocity semi-amplitudes of the primary and secondary components to be 
$K_1$ = 26.8$\pm$0.5 km s$^{-1}$ and $K_2$ = 203$\pm$3 km s$^{-1}$, respectively; hence, a mass ratio of $q$ = 0.132$\pm$0.003 
was determined. With these values and the inclination angle taken from Slawson et al. (2011), they obtained a semimajor axis 
of $a$ = 5.81$\pm$0.08 R$_\odot$ and masses of $M_1$ = 1.50$\pm$0.05 M$_\odot$ and $M_2$ = 0.198$\pm$0.007 M$_\odot$. 

In this article, we analyse in detail the {\it Kepler} photometric data of KIC 6206751 together with MGGW's RVs. 
From this analysis, we derive the unique physical properties of the binary system, and show that it is an R CMa-type EB 
with a $\gamma$ Dor-type pulsating component. In Section 2, the observations are discussed, including basic data reduction. 
In section 3, we present the binary modeling and absolute dimension. Section 4 describes the multiple frequency analyses 
for the residual light curve after removal of the binarity effects. We summarize and discuss our conclusions in section 5.

\section{OBSERVATIONS AND BASIC DATA REDUCTION}

KIC 6206751 was observed in both the long-cadence (LC) mode of 29.42 min and the short-cadence (SC) mode of 58.85 s by 
the {\it Kepler} satellite from 2009 to 2013 (Pr\v sa et al. 2011; Kirk et al. 2016). The {\it Kepler} data were collected 
in quarters (hereafter Q), approximately three months long. The ultra-precise photometric observations of the binary star 
were obtained in the LC mode during Q0$-$17 ($\sim$1470 d) and in the SC mode during Q4 ($\sim$20 d), Q11 ($\sim$97 d), and 
Q17 ($\sim$22 d). For this study, we used the simple aperture photometry data available at the Mikulski Archive for Space 
Telescopes (MAST)\footnote{http://archive.stsci.edu/kepler/}. The detrending process applied to the raw data was the same 
as that used by Lee et al. (2017), and the flux measurements were converted to a magnitude scale by requiring a {\it Kepler} 
magnitude of $+$12.142 at maximum light. The resultant {\it Kepler} data are shown in the top panel of Figure 1 as magnitudes 
versus the orbital phase. As well as the {\it Kepler} light curve, KIC 6206751 has the double-lined RV curves, which are 
obtained by MGGW and plotted in Figure 2. Because the times of the RV data are given as HJD, we transformed them into 
BJD$_{\rm TDB}$ using the online applets\footnote{http://astroutils.astronomy.ohio-state.edu/time/} of Eastman et al. (2010) 
so as to use the same timescale with the {\it Kepler} observations.

\section{BINARY MODELING AND ABSOLUTE DIMENSIONS}

As shown in Figure 1, the binary light curves of KIC 6206751 resemble those of classical Algols, and their shapes indicate 
an ellipsoidal variation and a considerable difference in the components' temperatures. Further, the scatter larger than 
about 0.03 mag is clearly seen in the light curves, which may be primarily produced by stellar activity such as oscillations. 
To obtain a consistent set of binary parameters, the {\it Kepler} light curve was solved with the RV measurements of MGGW 
in a manner similar to that for the pulsating EB OO Dra (Lee et al. 2018). We used the 2007 version of the Wilson-Devinney 
synthesis code (Wilson \& Devinney 1971, van Hamme \& Wilson 2007; hereafter W-D). Here, the subscripts 1 and 2 refer to 
the primary and secondary components being eclipsed at orbital phases 0.0 (Min I) and 0.5 (Min II), respectively.

In this synthesis, the effective temperature of the hotter primary component was set to be $T_1$ = 6965 K from 
the {\it Kepler} Input Catalogue (KIC; Kepler Mission Team 2009), which was used to build synthetic templates for 
cross-correlation with the observed spectra to measure RVs in MGGW. We assumed that the primary's temperature has an error 
of about 200 K that is the difference between the KIC and spectrosopic temperatures for dwarfs (Pinsonneault et al. 2012). 
The logarithmic bolometric ($X$, $Y$) and monochromatic ($x$, $y$) limb-darkening coefficients were interpolated from 
the values of van Hamme (1993). The gravity-darkening exponents were fixed at standard values of $g_1$ = 1.0 and $g_2$ = 0.32, 
while the bolometric albedos at $A_1$ = 1.0 and $A_2$ = 0.5. Furthermore, a synchronous rotation ($F_{1,2}$ = 1.0) 
for both components was adopted and the detailed reflection treatment was applied. The adjustable parameters became 
the orbital ephemeris ($T_0$, $P$, and d$P$/d$t$), the system velocity ($\gamma$), the semi-major axis ($a$), the mass ratio 
($q$), the orbital inclination ($i$), the surface temperature ($T_2$) of the secondary component, the dimensionless surface 
potentials ($\Omega_{1,2}$) of the components, the monochromatic luminosity ($L_{1}$), and the third light ($l_3$).

We separately analysed the LC and SC data, and the results are listed in the second to fifth columns of Table 1. For more 
reliable error estimates, we individually fitted the observed data of each quarter. The parameter errors presented in 
Table 1 are the standard deviation of each parameter computed from this procedure. The synthetic light curves are presented 
as solid curves in Figure 1, while the synthetic RV curves are displayed in Figure 2. On the other hand, the light curve 
parameters for the observed data could be affected by the multiperiodic pulsations presented in the following section. 
We analysed the pulsation-subtracted data after removing the pulsations from the observed SC data. Final result is displayed 
in Figure 3 and is given in columns (6) and (7) of Table 1. As listed in the table, the binary parameters from the observed 
and pulsation-subtracted data are consistent with each other within their errors. The light and RV solutions represent that 
the eclipsing pair is in a semi-detached configuration, where the primary component is slightly larger than 
the lobe-filling secondary and fills its limiting lobe by $\Omega_{\rm in}$/$\Omega_1$ $\simeq$ 50\%. Here, $\Omega_{\rm in}$ 
is the potential in the inner Lagrangian point. In all the procedures, we included the orbital eccentricity ($e$) as 
an additional free parameter, but its value remained zero. This indicates that KIC 6206751 is in a circular orbit, as 
expected for short-period semi-detached binaries. 

In Figures 1$-$3, our binary modeling provided a good fit to the light and RV curves, and it allowed us to determine 
the absolute dimensions of KIC 6206751. They are given at the bottom of Table 1, where the radii are the mean volume radii 
computed from the tables of Mochnacki (1984). The luminosity ($L$) and bolometric magnitudes ($M_{\rm bol}$) were derived 
by adopting $T_{\rm eff}$$_\odot$ = 5780 K and $M_{\rm bol}$$_\odot$ = +4.73 for solar values. The bolometric corrections 
(BCs) were obtained from the empirical relation between $\log T_{\rm eff}$ and BC given by Torres (2010). With an apparent 
visual magnitude of $V_{\max}$ = +12.055 taken from the ASAS catalog (Pigulski et al. 2009) and the colour excess of 
$E(B-V)$ = $+$0.188 (Kepler Mission Team 2009), we determined the distance of the system to be 555 $\pm$ 41 pc. This value 
is in satisfactory agreement with 452$\pm$72 pc computed from Gaia DR1 (2.21$\pm$0.35 mas; Gaia Collaboration et al. 2016).

\section{LIGHT CURVE RESIDUALS AND PULSATIONAL CHARACTERISTICS}

Considering the binary parameters in Table 1, the primary component of KIC 6206751 is a main-sequence star with a spectral 
type of about F0V located in the lower part of the Cepheid instability strip of the Hertzsprung-Russell (HR) diagram, 
which will be discussed in the following section. Then, the primary component would be a candidate for $\delta$ Sct and/or 
$\gamma$ Dor pulsations. We use all SC data to perform a frequency analysis of the light curve of the EB. The {\it Kepler} 
SC data were divided into 115 light curves at intervals of one orbital period and analysed them separately with the W-D code 
by adjusting only the ephemeris epoch ($T_0$) in Table 1. As an example, the light curve residuals for Q11 are displayed in 
Figure 4 as magnitudes versus BJDs, where the lower panel presents a short section of the outside-eclipse residuals. 

Using the discrete Fourier transform program PERIOD04 (Lenz \& Breger 2005), we applied multiple frequency analyses to 
all of the out-of-eclipse residuals of the SC data, which were performed up to the Nyquist limit of 
$f_{\rm Ny}$ = 732 d$^{-1}$. The periodogram from the PERIOD04 for the SC data is plotted in the top panel of Figure 5, 
where all the main signals have frequencies below 5 d$^{-1}$. As in our previous papers (Lee et al. 2014, 2016a,b, 2017; 
Lee 2016), a successive prewhitening of the data for all frequency contributions found in each step of analysis was 
carried out. This process was repeated until no significant frequency peaks were found. As a result, we detected 
at least 42 frequencies with signal-to-noise amplitude ratios (S/N) larger than 4.0 (Breger et al. 1993), and they 
are listed in Table 2. The uncertainties in the table were calculated according to Kallinger, Reegen \& Weiss (2008). 
The calculated frequency errors are much smaller than the Rayleigh frequency resolution of 1/$\Delta T$ = 0.00084 d$^{-1}$, 
where $\Delta T$ is the time base of observations. Thus, we estimated the uncertainties on the frequencies as 1/(4$\Delta T$) 
= 0.00021 d$^{-1}$ (Kallinger, Reegen \& Weiss 2008). The amplitude spectra after prewhitening the first four frequencies 
and then all 42 frequencies are presented in the second and third panels of Figure 5, respectively. The synthetic curve 
obtained from the 42-frequency fit is displayed as a solid line in the lower panel of Figure 4. Here, the blue symbols 
represent the residuals after removing the pulsations from the eclipse-subtracted data. 

Within the frequency resolution of 1.5/$\Delta T$ = 0.0012 d$^{-1}$ calculated according to Loumos \& Deeming (1978), 
we checked for the possible harmonic and combination frequencies. The results are given in the last column of Table 2. 
The $f_5$, $f_{12}$, and possible $f_1$ frequencies appear to be the orbital frequency ($f_{\rm orb}$ = 0.80299 d$^{-1}$) 
and its multiple, which could be partly affected by imperfect removal of the binary effects from the observed data. 
KIC 6206751 pulsates in three independent frequencies of $f_2$, $f_3$, and $f_4$ near 1.2 d$^{-1}$. On the other hand, 
we analysed all outside-eclipse LC residuals in the same manner as that applied to the SC data, and plotted a periodogram 
from the PERIOD04 program in the bottom panel of Figure 5. As a result of the pre-whitening for the LC data, we found 
three pulsation frequencies of 1.16287 d$^{-1}$, 1.23082 d$^{-1}$, and 1.18783 d$^{-1}$ in order of detection. 
The frequencies were almost identical to those from the {\it Kepler} SC data.

\section{DISCUSSION AND CONCLUSIONS}

In this paper, we presented and analysed the {\it Kepler} light curve of KIC 6206751 obtained during a four year period, 
together with the RV curves of MGGW. The binary modeling shows that the EB system is a semi-detached Algol with parameters 
of $q$ = 0.129, $i$ = 75$^\circ$.22, $T_1$ = 6965 K, $T_2$ = 4959 K, and $l_{3}$ = 5.81 \%, in which the detached primary 
fills about 50 \% of its inner Roche lobe. Because the contamination level of the {\it Kepler} observations by nearby stars 
is estimated to be 0.009, the third light source could be a circumbinary object gravitationally bound to the eclipsing pair, 
as suggested by the timing analysis (Gies et al. 2012, 2015). From the combined light and RV solution, 
the absolute dimensions for both components were determined to be $M_1$ = 1.66 M$_\odot$, $M_2$ = 0.215 M$_\odot$, 
$R_1$ = 1.53 R$_\odot$, $R_2$ = 1.33 R$_\odot$, $L_1$ = 5.0 L$_\odot$, and $L_2$ = 0.96 L$_\odot$. The short $P$, and 
low $q$ and $M_2$ of KIC 6206751 imply that it is an R CMa-type EB (Budding \& Butland 2011; Lehmann et al. 2013; 
Lee et al. 2016b). The program target may have formed via non-conservative binary evolution and most likely evolve into 
an EL CVn star (Chen et al. 2017; Lee et al. 2018).

The position of the components of KIC 6206751 in the HR diagram are displayed in Figure 6, together with those of 
the components of six well-studied R CMa-type stars and three EL CVn stars (cf. Lee et al. 2018). Here, the up- and 
down-pointing triangles denote detached and semi-detached configurations, respectively. The dashed and dash-dotted lines 
represent the theoretical blue and red edges of $\gamma$ Dor and $\delta$ Sct instability strips, respectively, and 
the cross symbols are $\gamma$ Dor pulsators in EBs with known parameters 
(Ibanoglu, \c Cakirli \& Sipahi 2018). The same figure displays three evolutionary models of low-mass white dwarfs with 
helium cores in the mass range of 0.179 M$_\odot$ to 0.234 M$_\odot$ calculated by Driebe et al. (1998). As shown in 
the figure, the primary component of KIC 6206751 lies in the $\gamma$ Dor region on the zero-age main sequence (ZAMS), and 
the low-mass secondary is noticeably evolved. As the result of non-conservative mass transfer (Chen et al. 2017), 
the original more massive star of KIC 6206751 became the present low-mass and oversized secondary component, and the gainer 
became the pulsating primary located in the $\gamma$ Dor region due to mass accretion. Later on, our program target will 
become a detached R CMa-type star similar to KIC 8262223 (Guo et al. 2017) and OO Dra (Lee et al. 2018) after stopping 
its mass transfer. Then, KIC 6206751 will evolve into an EL CVn-like star, following the evolutionary tracks presented in 
Figure 6. 

To find the pulsation frequencies of KIC 6206751, we removed the binarity effects from the observed {\it Kepler} SC data 
and performed a multiple frequency analysis in the whole outside-eclipse light residuals. As a consequence, we detected 
42 frequencies with S/N values larger than the empirical threshold of 4.0. Three ($f_2$, $f_3$, and $f_4$) of these may 
be pulsation frequencies originating from the primary component, while the other frequencies may be orbital harmonics and 
combination terms. Applying the absolute parameters in the pulsation-subtracted SC data of Table 1 to the equation of 
$\log Q_i = -\log f_i + 0.5 \log g + 0.1M_{\rm bol} + \log T_{\rm eff} - 6.456$ (Petersen \& J\o rgensen 1972), we obtained 
the pulsation constants $Q$ as listed in Table 3. The pulsation periods ($P_{\rm pul}$) of 0.812$-$0.859 d and the $Q$ values 
of 0.550$-$0.582 d correspond to the $g$ modes of typical $\gamma$ Dor pulsators (Kaye et al. 1999; Henry, Fekel \& Henry 2005). 
The ratios of the orbital frequency to the three pulsation frequencies are $f_{\rm orb}$:$f_{\rm 2-4}$ $\simeq$ 2:3. 
The results indicate that the $\gamma$ Dor pulsations of KIC 6206751 may be excited by the tidal interaction of 
the secondary companion onto the pulsating primary. 

If a $\gamma$ Dor star has a rotational velocity of $v \sin i \la$ 70 km s$^{-1}$ and at least three $g$-mode frequencies, 
it is possible to identify the radial order ($n$) and spherical degree ($\ell$) for the pulsation frequencies with 
the Frequency Ratio Method (Moya et al. 2005; Su\'arez et al. 2005). The rotational velocity of KIC 6206751 is not known, 
but it is a short-period semi-detached EB; hence, it is expected to be in a synchronised rotation. Then, the pulsating 
primary may have a synchronous rotation of $v_{\rm sync}$ $\simeq$ 62 km s$^{-1}$. The pulsation modes listed in Table 3 
for the three independent frequencies were identified following the procedure used by Lee et al. (2014). We found 
the model frequency ratios ($f_i$/$f_2$)$_{\rm model}$ best-fitting to the observed ratios ($f_i$/$f_2$)$_{\rm obs}$. 
The $f_2$, $f_3$, and $f_4$ frequencies were identified as degree $\ell$ = 2 for three consecutive radial orders of 
$n$ = 38, 39, and 40, respectively. The observed Brunt-V\"ais\"al\"a frequency integral of $\cal J_{\rm obs}$ = 700.1$\pm$3.5 
is a good match to the theoretical value of $\cal J_{\rm theo}$ $\approx$ 700 $\mu$Hz for a model of 
$\log$ $T_{\rm eff}$ = 3.843, 1.6 M$_\odot$, and [Fe/H] = 0.0 in the $\cal J -$ $\log$ $T_{\rm eff}$ diagram described in 
Figure 5 of Moya et al. (2005). Including our program target, seven EBs with reliable physical properties are known to be 
R CMa-like stars, of which KIC 6206751 is the first binary system displaying $\gamma$ Dor pulsations. The results presented 
in this paper make KIC 6206751 an ideal target for asteroseismology and the study of EL CVn stars.

\section*{Acknowledgments}
We appreciate the careful reading and valuable comments of the anonymous referee. This paper includes data collected by 
the {\it Kepler} mission. {\it Kepler} was selected as the 10th mission of the Discovery Program. Funding for the {\it Kepler} 
mission is provided by the NASA Science Mission directorate. We have used the Simbad database maintained at CDS, Strasbourg, 
France. This work was supported by the KASI grant 2018-1-830-02.

\clearpage
\begin{figure}
\includegraphics{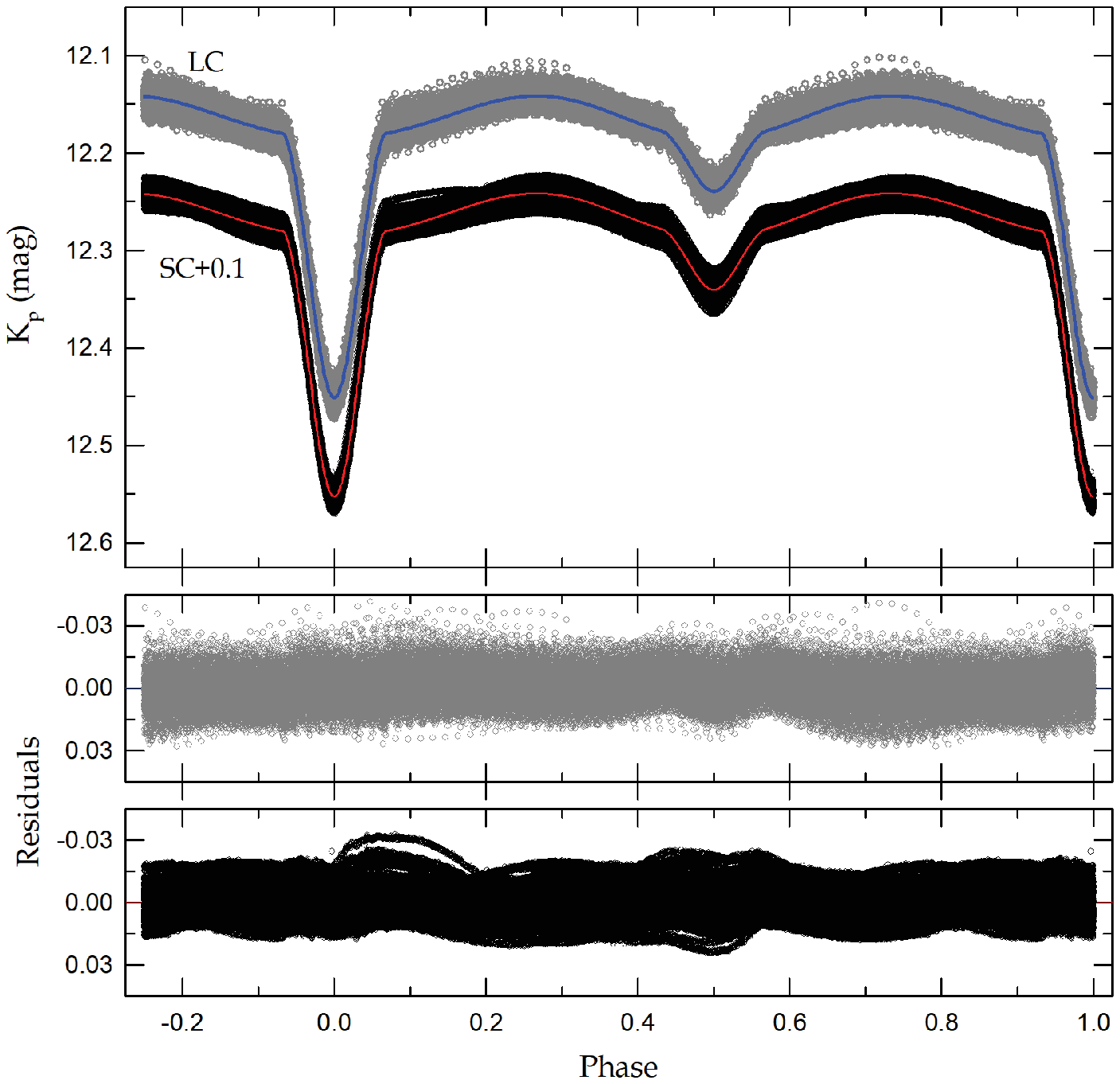}
\caption{Binary light curve of KIC 6206751 with the fitted models. The gray and black circles are the LC and SC data 
observed from the {\it Kepler} spacecraft, respectively, and the blue and red curves represent the synthetic curves for 
the two datasets in Table 1. The corresponding residuals from the fits are plotted at the middle and bottom panels in 
the same order as the light curves. }
\label{Fig1}
\end{figure}

\begin{figure}
\includegraphics[]{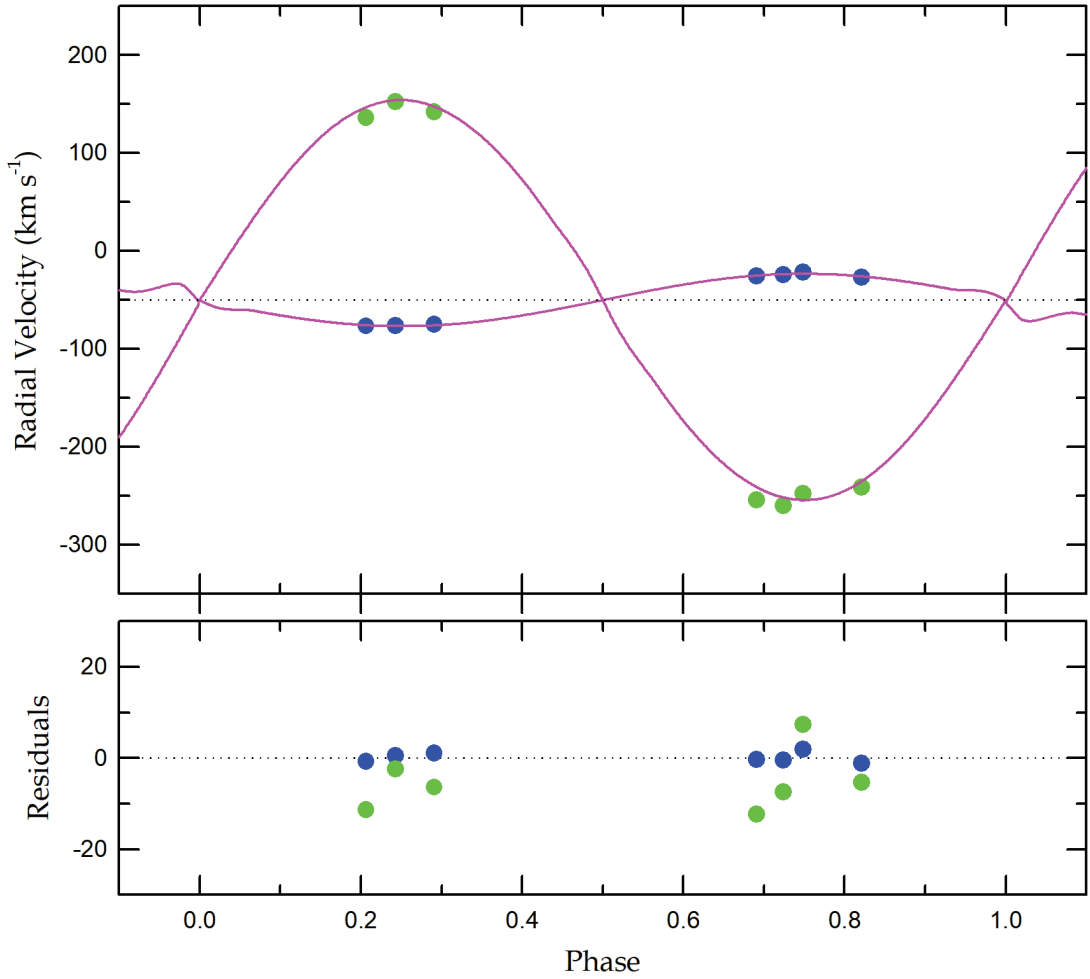}
\caption{Radial-velocity curves of KIC 6206751. The blue and green circles are the primary and secondary measurements of 
Matson et al. (2017), respectively. In the upper panel, the solid curves represent the results from a consistent light and 
RV curve analysis including proximity effects. The dotted line refers to the system velocity of $-$50.4 km s$^{-1}$. 
The lower panel shows the residuals between observations and models. }
\label{Fig2}
\end{figure}

\begin{figure}
\includegraphics{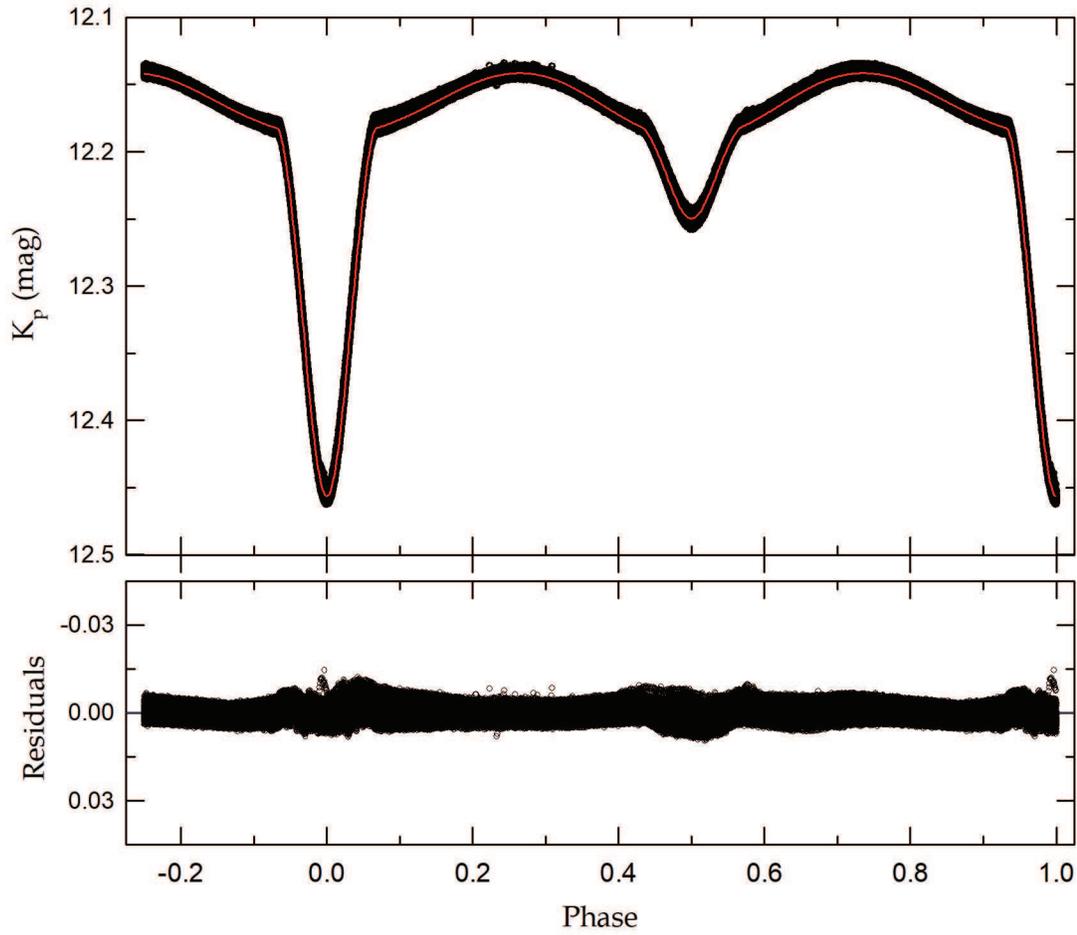}
\caption{Binary light curve after subtracting the pulsation signatures from the observed SC data. The solid curve in 
the upper panel was obtained from the pulsation-subtracted data using the W-D code and the corresponding residuals are 
plotted in the lower panel. Some features visible during the times of both eclipses may come from insufficient removal of 
the pulsations in the orbital phases, which is caused by using only outside-eclipse data in the frequency analysis. }
\label{Fig3}
\end{figure}

\begin{figure}
\includegraphics[]{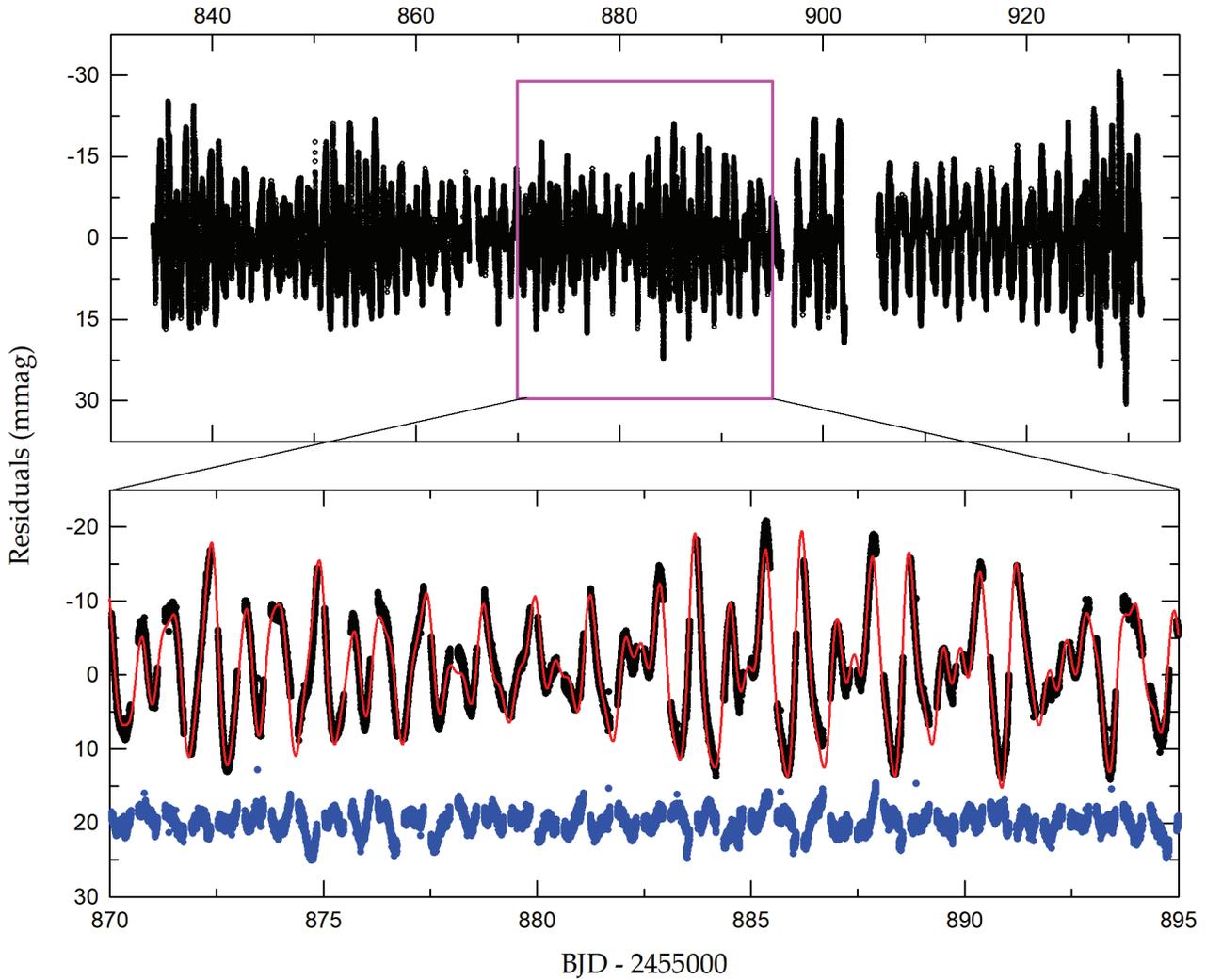}
\caption{Light curve residuals for the SC data of Q11. The lower panel presents a short section of the residuals marked 
using the inset box of the upper panel. The synthetic curve is computed from the 42-frequency fit to the out-of-eclipse 
data. The residuals from the fit are offset from zero by +20 mmag for clarity and are plotted as the blue symbols. }
\label{Fig4}
\end{figure}

\begin{figure}
\includegraphics[]{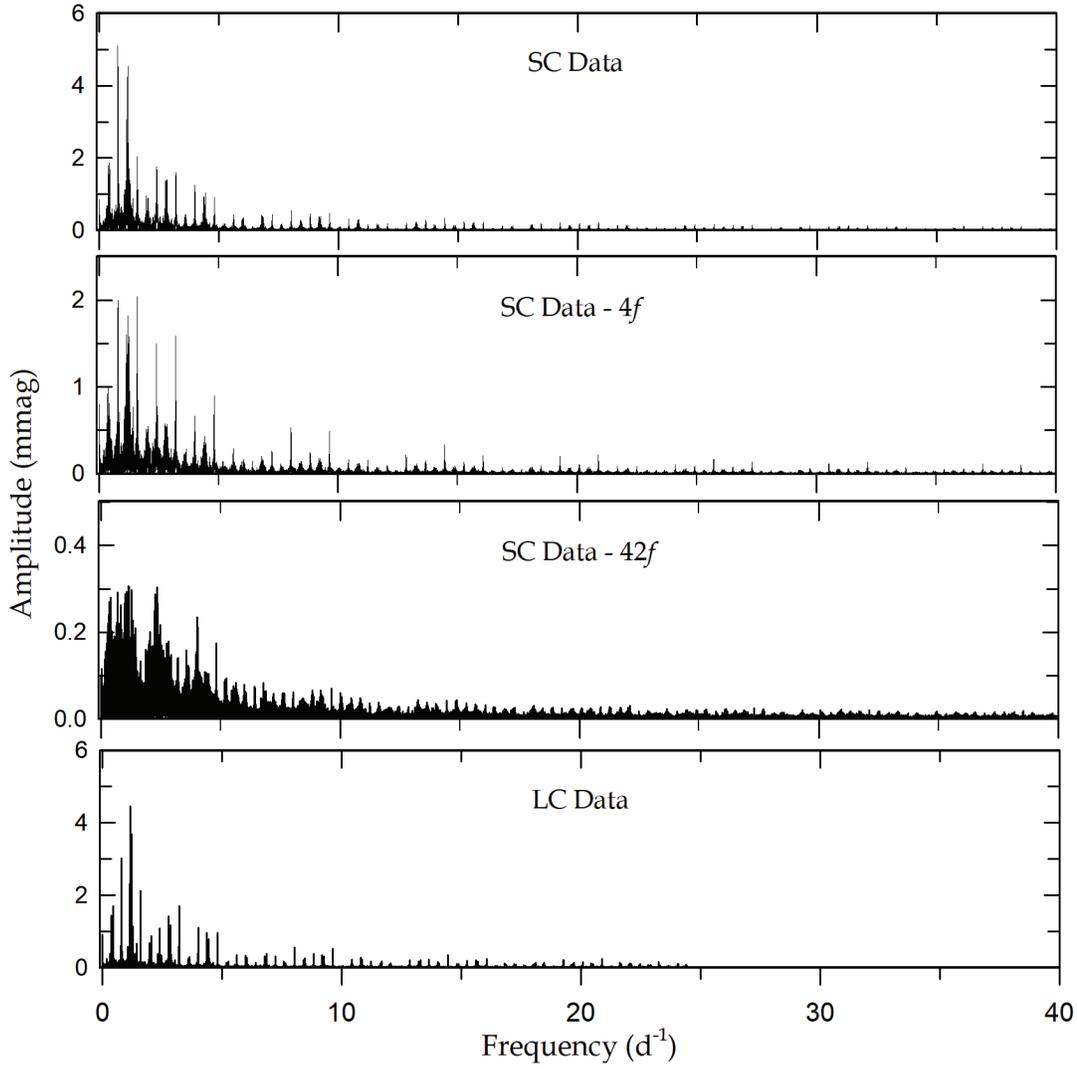}
\caption{Amplitude spectra before (top panel) and after pre-whitening the first four frequencies (second) and all 42 
frequencies (third) from the PERIOD04 program for all SC light residuals. The periodogram for the LC data is shown in 
the bottom panel for comparison. }
\label{Fig5}
\end{figure}

\begin{figure}
\includegraphics[]{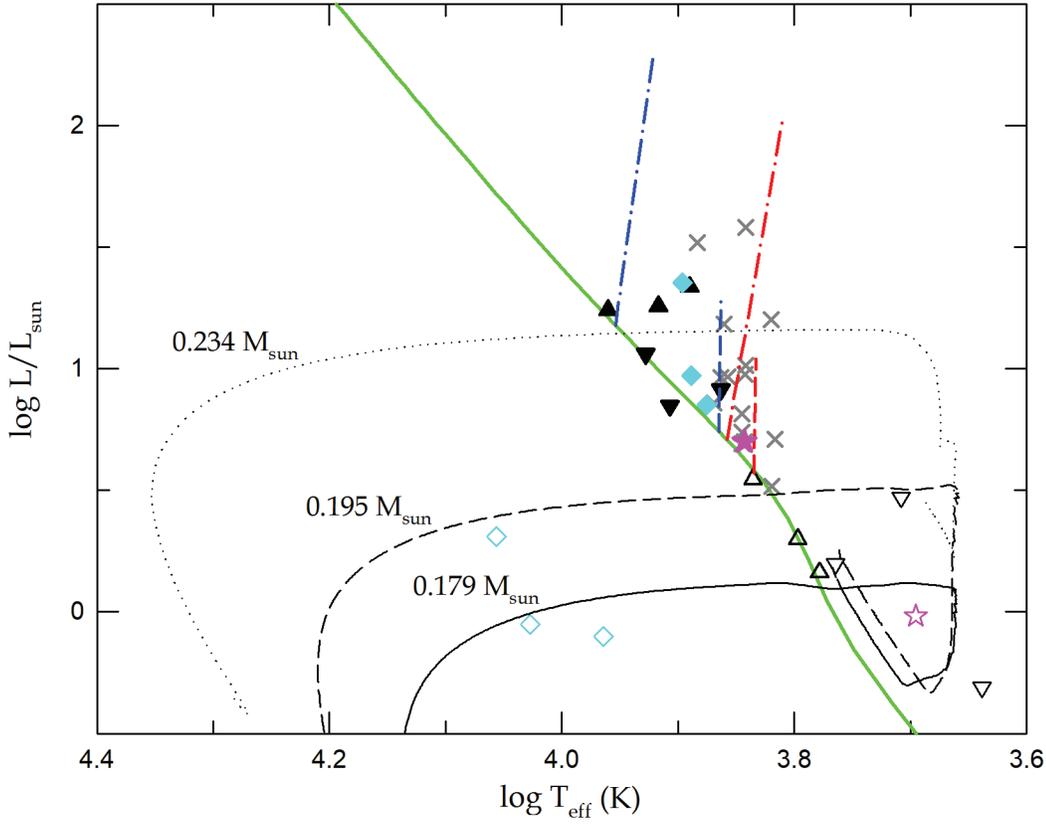}
\caption{Positions in the HR diagrams of KIC 6206751 (star symbols), $\gamma$ Dor pulsators in EBs (crosses), and R CMa 
(triangles) and EL CVn (diamonds) stars. The filled and open symbols refer to the primary and secondary components, 
respectively. The solid green line represents the ZAMS for solar metallicity from Tout et al. (1996). The coloured dashed and 
dash-dotted lines represent the instability strips of $\gamma$ Dor (Warner, Kaye \& Guzik 2003; \c Cakirli 2015) and 
$\delta$ Sct stars (Rolland et al. 2002; Soydugan et al. 2006), respectively. The solid, dashed, and dotted black lines 
represent the evolutionary track of helium white dwarfs with masses of 0.179 M$_\odot$, 0.195 M$_\odot$, and 0.234 M$_\odot$ 
(Driebe et al. 1998), respectively. }
\label{Fig6}
\end{figure}

\clearpage
\begin{table}
\caption{Binary parameters of KIC 6206751. Errors in units of the last digits are given in parentheses. }
\begin{tabular}{lcccccccc}
\hline
Parameter                                & \multicolumn{2}{c}{LC Data$^a$}                  && \multicolumn{2}{c}{SC Data$^a$}                  && \multicolumn{2}{c}{SC Data$^b$}                    \\ [1.0mm] \cline{2-3} \cline{5-6} \cline{8-9} \\[-2.0ex]
                                         & Primary           & Secondary                    && Primary           & Secondary                    && Primary           & Secondary                      \\                                                                                         
\hline                                                                                                                                                                                              
$T_0$ (BJD)                              & \multicolumn{2}{c}{2,455,690.00953(14)}          && \multicolumn{2}{c}{2,455,690.00954(10)}          && \multicolumn{2}{c}{2,455,690.00996(3)}             \\ 
$P$ (day)                                & \multicolumn{2}{c}{1.24534194(29)}               && \multicolumn{2}{c}{1.24534324(30)}               && \multicolumn{2}{c}{1.24534258(9)}                  \\
d$P$/d$t$                                & \multicolumn{2}{c}{$-$7.2(1.5)$\times$10$^{-9}$} && \multicolumn{2}{c}{$-$6.4(1.2)$\times$10$^{-9}$} && \multicolumn{2}{c}{$-$7.3(3)$\times$10$^{-9}$}     \\
$a$ (R$_\odot$)                          & \multicolumn{2}{c}{6.009(64)}                    && \multicolumn{2}{c}{6.005(63)}                    && \multicolumn{2}{c}{6.005(19)}                      \\
$\gamma$ (km s$^{-1}$)                   & \multicolumn{2}{c}{$-$50.50(37)}                 && \multicolumn{2}{c}{$-$50.40(37)}                 && \multicolumn{2}{c}{$-$50.40(11)}                   \\
$q$                                      & \multicolumn{2}{c}{0.1295(31)}                   && \multicolumn{2}{c}{0.1294(30)}                   && \multicolumn{2}{c}{0.1294(22)}                     \\
$i$ (deg)                                & \multicolumn{2}{c}{75.09(20)}                    && \multicolumn{2}{c}{75.22(42)}                    && \multicolumn{2}{c}{75.22(1)}                       \\
$T$ (K)                                  & 6965(200)         & 4855(102)                    && 6965(200)         & 4864(104)                    && 6965(200)         & 4959(110)                      \\
$\Omega$                                 & 4.039(26)         & 2.046                        && 4.065(33)         & 2.046                        && 4.064(2)          & 2.046                          \\
$\Omega_{\rm in}$                        & \multicolumn{2}{c}{2.046}                        && \multicolumn{2}{c}{2.046}                        && \multicolumn{2}{c}{2.046}                          \\
$A$                                      & 1.0               & 0.5                          && 1.0               & 0.5                          && 1.0               & 0.5                            \\
$g$                                      & 1.0               & 0.32                         && 1.0               & 0.32                         && 1.0               & 0.32                           \\
$X$, $Y$                                 & 0.640, 0.253      & 0.642, 0.160                 && 0.640, 0.253      & 0.642, 0.160                 && 0.640, 0.253      & 0.643, 0.166                   \\
$x$, $y$                                 & 0.615, 0.294      & 0.742, 0.155                 && 0.615, 0.294      & 0.742, 0.156                 && 0.615, 0.294      & 0.737, 0.171                   \\
$l$/($l_{1}$+$l_{2}$+$l_{3}$)            & 0.7965(34)        & 0.1370                       && 0.7891(42)        & 0.1387                       && 0.7892(9)         & 0.1527                         \\
$l_{3}$$\rm ^c$                          & \multicolumn{2}{c}{0.0665(40)}                   && \multicolumn{2}{c}{0.0721(31)}                   && \multicolumn{2}{c}{0.0581(9)}                      \\
$r$ (pole)                               & 0.2555(18)        & 0.2051(15)                   && 0.2539(22)        & 0.2051(15)                   && 0.2539(19)        & 0.2051(14)                     \\
$r$ (point)                              & 0.2590(19)        & 0.3036(20)                   && 0.2573(24)        & 0.3035(20)                   && 0.2573(23)        & 0.3035(19)                     \\
$r$ (side)                               & 0.2580(18)        & 0.2133(15)                   && 0.2563(23)        & 0.2132(15)                   && 0.2563(20)        & 0.2132(15)                     \\
$r$ (back)                               & 0.2587(19)        & 0.2448(16)                   && 0.2570(23)        & 0.2448(16)                   && 0.2570(22)        & 0.2448(15)                     \\
$r$ (volume)$\rm ^d$                     & 0.2574(18)        & 0.2215(16)                   && 0.2557(23)        & 0.2214(16)                   && 0.2558(21)        & 0.2214(16)                     \\
$\sum W(O-C)^2$                          & \multicolumn{2}{c}{0.0072}                       && \multicolumn{2}{c}{0.0070}                       && \multicolumn{2}{c}{0.0021}                         \\[1.0mm]
\multicolumn{6}{l}{Absolute parameters:}                                                                                                                                                              \\            
$M$ (M$_\odot$)                          & 1.66(4)           & 0.216(7)                     && 1.66(4)           & 0.215(7)                     && 1.66(4)           & 0.215(6)                       \\
$R$ (R$_\odot$)                          & 1.55(2)           & 1.33(2)                      && 1.53(2)           & 1.33(2)                      && 1.53(2)           & 1.33(2)                        \\
$\log$ $g$ (cgs)                         & 4.28(2)           & 3.52(2)                      && 4.29(2)           & 3.52(2)                      && 4.29(1)           & 3.52(2)                        \\
$L$ (L$_\odot$)                          & 5.0(6)            & 0.88(8)                      && 5.0(6)            & 0.89(8)                      && 5.0(6)            & 0.96(9)                        \\
$M_{\rm bol}$ (mag)                      & 2.97(13)          & 4.87(10)                     && 2.99(13)          & 4.86(10)                     && 2.99(13)          & 4.78(10)                       \\
BC (mag)                                 & 0.03(1)           & $-$0.37(6)                   && 0.03(1)           & $-$0.37(6)                   && 0.03(1)           & $-$0.31(5)                     \\
$M_{\rm V}$ (mag)                        & 2.94(13)          & 5.24(11)                     && 2.96(13)          & 5.23(12)                     && 2.96(13)          & 5.09(11)                       \\
Distance (pc)                            & \multicolumn{2}{c}{556$\pm$41}                   && \multicolumn{2}{c}{554$\pm$41}                   && \multicolumn{2}{c}{555$\pm$41}                     \\
\hline
\multicolumn{9}{l}{$^a$ Result from the observed data.} \\
\multicolumn{9}{l}{$^b$ Result from the pulsation-subtracted data.} \\
\multicolumn{9}{l}{$^c$ Value at 0.25 phase.} \\
\multicolumn{9}{l}{$^d$ Mean volume radius.} 
\end{tabular}
\end{table}

\begin{table}
\caption{Frequencies found in the SC light curve residuals of KIC 6206751. Possible combination frequencies are indicated in the last column. }
\begin{tabular}{lccccc}
\hline
             & Frequency              & Amplitude           & Phase           & S/N            & Remark                     \\
             & (day$^{-1}$)           & (mmag)              & (rad)           &                &                            \\
\hline
$f_{1}$      & 0.79307$\pm$0.00001    & 4.76$\pm$0.14       & 3.04$\pm$0.08   & 60.02          & $f_{\rm orb}$(?)           \\
$f_{2}$      & 1.23091$\pm$0.00001    & 4.17$\pm$0.13       & 4.23$\pm$0.09   & 54.75          &                            \\
$f_{3}$      & 1.18796$\pm$0.00001    & 3.68$\pm$0.13       & 5.26$\pm$0.10   & 47.87          &                            \\
$f_{4}$      & 1.16461$\pm$0.00001    & 3.02$\pm$0.13       & 2.42$\pm$0.13   & 39.25          &                            \\
$f_{5}$      & 1.60613$\pm$0.00001    & 4.69$\pm$0.12       & 2.13$\pm$0.08   & 64.98          & $2f_{\rm orb}$             \\
$f_{6}$      & 0.80801$\pm$0.00001    & 2.10$\pm$0.14       & 3.33$\pm$0.19   & 26.52          & $f_2+f_4-2f_1$             \\
$f_{7}$      & 1.22273$\pm$0.00001    & 2.09$\pm$0.13       & 0.10$\pm$0.18   & 27.27          & $f_1+2f_6-f_3$             \\
$f_{8}$      & 1.15679$\pm$0.00001    & 2.02$\pm$0.13       & 2.37$\pm$0.19   & 26.24          & $f_4+f_7-f_2$              \\
$f_{9}$      & 0.79987$\pm$0.00001    & 1.81$\pm$0.14       & 1.58$\pm$0.22   & 22.82          & $f_1+f_4-f_8$              \\
$f_{10}$     & 1.19951$\pm$0.00001    & 1.50$\pm$0.13       & 1.94$\pm$0.26   & 19.46          & $f_2+f_8-f_3$              \\
$f_{11}$     & 1.26590$\pm$0.00002    & 1.35$\pm$0.13       & 0.91$\pm$0.28   & 17.77          & $f_2+f_7-f_3$              \\
$f_{12}$     & 1.60480$\pm$0.00001    & 2.25$\pm$0.12       & 1.72$\pm$0.16   & 31.15          & $2f_{\rm orb}$             \\
$f_{13}$     & 1.17569$\pm$0.00002    & 1.25$\pm$0.13       & 2.28$\pm$0.31   & 16.26          & $f_1+f_5-f_7$              \\
$f_{14}$     & 2.41320$\pm$0.00001    & 1.41$\pm$0.11       & 4.47$\pm$0.23   & 22.15          & $f_5+f_6$                  \\
$f_{15}$     & 0.78605$\pm$0.00002    & 1.35$\pm$0.14       & 2.63$\pm$0.30   & 16.97          & $f_1+f_7-f_2$              \\
$f_{16}$     & 0.43958$\pm$0.00003    & 0.86$\pm$0.14       & 4.81$\pm$0.47   & 10.65          & $f_{12}-f_4$               \\
$f_{17}$     & 0.37277$\pm$0.00002    & 0.96$\pm$0.14       & 1.17$\pm$0.42   & 11.86          & $f_{14}-f_1$               \\
$f_{18}$     & 1.13952$\pm$0.00002    & 1.26$\pm$0.13       & 5.22$\pm$0.31   & 16.33          & $f_5+f_9-f_{11}$           \\
$f_{19}$     & 1.05997$\pm$0.00003    & 0.69$\pm$0.13       & 1.07$\pm$0.57   &  8.82          & $f_{18}+f{9}-2f_{16}$      \\
$f_{20}$     & 1.10735$\pm$0.00003    & 0.77$\pm$0.13       & 0.02$\pm$0.50   &  9.99          & $2f_4-f_7$                 \\
$f_{21}$     & 1.29310$\pm$0.00002    & 0.88$\pm$0.13       & 6.02$\pm$0.43   & 11.57          & $f3+f_4-f_{19}$            \\
$f_{22}$     & 1.24695$\pm$0.00002    & 0.91$\pm$0.13       & 4.76$\pm$0.42   & 11.92          & $f_{16}+f_6$               \\
$f_{23}$     & 1.42475$\pm$0.00003    & 0.79$\pm$0.13       & 1.52$\pm$0.47   & 10.70          & $f_{19}+f_8-f_1$           \\
$f_{24}$     & 0.41579$\pm$0.00003    & 0.72$\pm$0.14       & 1.74$\pm$0.56   &  8.91          & $f_{12}-f_3$               \\
$f_{25}$     & 2.40309$\pm$0.00002    & 1.00$\pm$0.11       & 2.19$\pm$0.32   & 15.61          & $f_{22}+f_8$               \\
$f_{26}$     & 0.77963$\pm$0.00003    & 0.81$\pm$0.14       & 2.69$\pm$0.49   & 10.21          & $f_1+f_{15}-f_9$           \\
$f_{27}$     & 1.20769$\pm$0.00003    & 0.68$\pm$0.13       & 5.99$\pm$0.56   &  8.89          & $f_1+f_{24}$               \\
$f_{28}$     & 2.45995$\pm$0.00004    & 0.46$\pm$0.11       & 3.50$\pm$0.69   &  7.23          & $f_{14}+f_7-f_{13}$        \\
$f_{29}$     & 1.36516$\pm$0.00004    & 0.54$\pm$0.13       & 2.42$\pm$0.69   &  7.29          & $f_{23}+f_3-f_{22}$        \\
$f_{30}$     & 1.32709$\pm$0.00004    & 0.53$\pm$0.13       & 2.71$\pm$0.71   &  7.06          & $f_2+f_8-f_{19}$           \\
$f_{31}$     & 0.38146$\pm$0.00004    & 0.54$\pm$0.14       & 4.49$\pm$0.75   &  6.65          & $f_{13}-f_1$               \\
$f_{32}$     & 1.61240$\pm$0.00003    & 0.68$\pm$0.12       & 4.37$\pm$0.53   &  9.47          & $f_{14}-f_9$               \\
$f_{33}$     & 1.12495$\pm$0.00004    & 0.52$\pm$0.13       & 3.07$\pm$0.74   &  6.78          & $2f_8-f_3$                 \\
$f_{34}$     & 1.07979$\pm$0.00005    & 0.46$\pm$0.13       & 1.06$\pm$0.84   &  5.96          & $f_3+f_8-f_{11}$           \\
$f_{35}$     & 1.26038$\pm$0.00004    & 0.53$\pm$0.13       & 3.16$\pm$0.72   &  6.96          & $f_{28}-f_{10}$            \\
$f_{36}$     & 2.32133$\pm$0.00005    & 0.39$\pm$0.11       & 5.80$\pm$0.81   &  6.15          & $f_4+f_8$                  \\
$f_{37}$     & 2.44287$\pm$0.00004    & 0.41$\pm$0.11       & 0.50$\pm$0.78   &  6.41          & $f_{14}+f_{35}-f_2$        \\
$f_{38}$     & 0.75863$\pm$0.00006    & 0.39$\pm$0.14       & 3.98$\pm$1.02   &  4.90          & $f_{18}-f_{31}$            \\
$f_{39}$     & 0.34699$\pm$0.00006    & 0.40$\pm$0.14       & 5.84$\pm$1.02   &  4.93          & $f_{18}-f_1$               \\
$f_{40}$     & 1.38588$\pm$0.00006    & 0.35$\pm$0.13       & 2.08$\pm$1.06   &  4.71          & $f_{30}+f_7-f_4$           \\
$f_{41}$     & 1.30733$\pm$0.00006    & 0.38$\pm$0.13       & 6.22$\pm$0.99   &  5.04          & $f_{21}+f_6-f_1$           \\
$f_{42}$     & 1.58966$\pm$0.00005    & 0.41$\pm$0.12       & 2.03$\pm$0.89   &  5.65          & $f_{27}+f_{31}$            \\
\hline
\end{tabular}
\end{table}

\begin{table}
\caption{$\gamma$ Dor-type Pulsation Properties of KIC 6206751. }
\begin{tabular}{lccccccc}
\hline
         & Frequency      & $Q$      & ($f_i$/$f_2$)$_{\rm obs}$   & mode ($n$, $\ell$)   & ($f_i$/$f_2$)$_{\rm model}$   & $\Delta$($f_i$/$f_2$)$_{\rm obs-model}$    & $\cal J_{\rm obs}$      \\
         & (day$^{-1}$)   & (days)   &                             &                      &                               &                                            & ($\mu$Hz)               \\
\hline
$f_{2}$  & 1.23091   & 0.550  & \dots    & (38, 2)    & \dots    & \dots                 & 703.5            \\
$f_{3}$  & 1.18796   & 0.570  & 0.9651   & (39, 2)    & 0.9747   & $-$0.0096             & 696.6            \\
$f_{4}$  & 1.16461   & 0.582  & 0.9461   & (40, 2)    & 0.9506   & $-$0.0045             & 700.2            \\
         &           &        &          &            & Average  & $-$0.0070$\pm$0.0036  & 700.1$\pm$3.5    \\  
\hline
\end{tabular}
\end{table}

\bsp
\label{lastpage}
\end{document}